\documentstyle[twocolumn,prd,aps]{revtex}

\begin{document}

\author{J.W.~Bos$^1$, D.W.~Chang$^2$, S.C.~Lee$^3$, Y.C.~Lin$^1$,
and H.H.~Shih$^1$}
\address{${}^1$Department of Physics and Astronomy, National
Central University, Chungli, Taiwan
\\
${}^2$Department of Physics, National Tsing Hua University, Hsinchu,
Taiwan
\\
${}^3$Institute of Physics, Academia Sinica, Taipei, Taiwan}

\title{Baryon chiral perturbation theory up to next-to-leading
order}
\date{\today}
\maketitle

\begin{abstract}
We examine the general lagrangian for baryon chiral perturbation theory
with SU(3) flavor symmetry, up to the next-to-leading order. We
consider both the strong and the weak interaction.  The inverse of the
baryon mass is treated as an additional small expansion parameter, and
heavy fermion effective field theory techniques are employed to provide
a consistent expansion scheme.  A detailed account is given on the
restrictions imposed on the lagrangian by the various symmetries.
Corrections due to the finite baryon mass are also discussed.
\end{abstract}

\pacs{11.30.RD}


\section{Introduction}

Chiral perturbation theory \cite{gass85} provides a promising way to
get insight in low-energy processes involving baryons.  In chiral
perturbation theory one starts with the  most general lagrangian in
terms of baryon and meson degrees of freedom, employing the chiral
symmetry of the underlying QCD lagrangian in the massless quark limit.
Based on this lagrangian, a perturbation scheme is then developed in
which one expands in the momenta of the external particles and
simultaneously in the mass of the Goldstone bosons.

However, compared to when it is applied to the meson sector
\cite{gass84}, chiral perturbation theory applied to the baryon sector
is more complicated. In the baryon sector, a loop expansion typically
will generate powers of $\dot{m}/\Lambda_{\chi\text{SB}}$, where
$\dot{m}$ is the nucleon mass in the chiral limit and
$\Lambda_{\chi\text{SB}}$ is the symmetry breaking scale of chiral
perturbation theory.  Since $\dot{m}/\Lambda_{\chi\text{SB}} \approx
1$, the expansion scheme seems to break down.  To avoid this
complication, it was suggested\cite{jenk91} to use heavy quark
effective field theory techniques---developed originally to treat heavy
quark systems---to reformulate baryon chiral perturbation theory with a
modified expansion scheme. This is done by redefining the effective
baryon field.  A loop expansion will now give rise to powers of
$k/\Lambda_{\chi\text{SB}}$, where $k$ is a small ``residual'' nucleon
four-momentum, making a systematic expansion feasible.  The formulation
consists of an simultaneous expansion in $k$, the mass of the strange
quark $m_{\text{s}}$ (for simplicity we neglect the up- and down quark
masses), and $1/\dot{m}$.  We will refer to this formulation as heavy
baryon chiral perturbation theory (HBCPT).

In this paper we study the HBCPT lagrangian, with SU(3) flavor
symmetry, for both the strong and weak interaction sector.  We put
emphasis on an essential feature of phenomenological lagrangians,
namely that one must include {\em all\/} terms consistent with the
assumed symmetry properties \cite{wein79}.

In the strong interaction sector, the leading order lagrangian is of
order ${\cal O}(k)$, i.e., independent of $m_{\text{s}}$, and
$\dot{m}$.  The next order, or {\em next-to-leading order}, corrections
consist of all terms of order ${\cal O}(k^2)$, ${\cal
O}(m_{\text{s}})$, and ${\cal O}(1/\dot{m})$.  Note that this
classification is only by convention: The relative counting between $k$
and $m_{\text s}$ is {\em a priori\/} not clear and can only be
determined by experiments.  In the weak interaction sector, the leading
order lagrangian is of order ${\cal O}(1)$, and the next-to-leading
order lagrangian consists by definition of all terms of order ${\cal
O}(k)$, ${\cal O}(m_{\text{s}})$, and ${\cal O}(1/\dot{m})$.

Loop diagrams generated with the leading order HBCPT lagrangian
contribute only beyond the next-to-leading order \cite{rho93}, i.e.,
the most important corrections to a leading-order amplitude come from
the next-to-leading order lagrangian at tree-level. This clearly makes
it necessary for a consistent calculation to take into account the full
structure of the next-to-leading order lagrangian.

This paper is organized as follows.  In Sec.~\ref{strong} we will
discuss the HBCPT lagrangian in the strong interaction sector.  First,
we will rewrite the leading order lagrangian using the heavy fermion
approach.  To obtain the next-to-leading order lagrangian we will
examine the conditions imposed on a general term by the various
symmetry requirements.  The possible linear relations between single
and double trace terms in the lagrangian is considered closely.  We
will then give the $1/\dot{m}$-lagrangian, using an approach based on
the equation of motion. In Sec.~\ref{weak} we look at the weak
interaction, and finally, Sec.~\ref{conclusions} contains a brief
summary and our conclusions.


\section{Strong interaction lagrangian}
\label{strong}


\subsection{Leading order}

The leading order strong interaction lagrangian of baryon chiral
perturbation theory is given by \cite{gass88,krau90}
\begin{eqnarray}
\label{lag1}
	{\cal L}&=&\text{Tr}\,\bar{B}i\gamma^\mu[D_\mu,B]-
	\dot{m}\bar{B}B
	+D\bar{B}i\gamma^5\gamma^\mu\{\Delta_\mu,B\}
\nonumber\\&&\mbox{}
	+F\bar{B}i\gamma^5\gamma^\mu[\Delta_\mu,B] \;,
\end{eqnarray}
where $B$ is the baryon field
\begin{equation}
	B=\left(\begin{array}{ccc}\frac{1}{\sqrt{6}}\Lambda+
	\frac{1}{\sqrt{2}}\Sigma^0 &
	\Sigma^+&p\\\Sigma^-&
	\frac{1}{\sqrt{6}}\Lambda-\frac{1}{\sqrt{2}}\Sigma^0 &n\\
	\Xi^-&\Xi^0&-\frac{2}{\sqrt{6}}\Lambda\end{array}\right) \;,
\end{equation}
and $\dot{m}$ is the nucleon mass in the chiral limit. The mesons are
contained in the anti-hermitian field
$\Delta^\mu$, given by
\begin{equation}
	\Delta^\mu
 	=\frac{1}{2}(\xi^{\dagger}\partial^\mu\xi-\xi\partial^\mu
	\xi^{\dagger}) \;,
\end{equation}
where $\xi$ is defined by $\xi^2=\Sigma$.  Here $\Sigma$ is the SU(3)
matrix
\begin{equation}
	\Sigma =\exp (2i\pi/f_{\pi}) \;,
\end{equation}
with
\begin{equation}
	\pi=\frac{1}{\sqrt{2}}
	\left(\begin{array}{ccc}\frac{1}{\sqrt{2}}\pi^0+
	\frac{1}{\sqrt{6}}\eta &\pi^+&K^+\\\pi^-&-
	\frac{1}{\sqrt{2}}\pi^0+
	\frac{1}{\sqrt{6}}\eta &K^0\\K^-&
	\overline{K^0}&-\frac{2}{\sqrt{6}}\eta\end{array}\right) \;
\end{equation}
and $f_{\pi}$ is the pion decay constant, $f_{\pi}=94 \text{ MeV}$.
The meson fields also appear in the covariant derivative $D_\mu$, given
by
\begin{equation}
	D^\mu=\partial^\mu+\Gamma^\mu \;,
\end{equation}
with $\Gamma^\mu$ the anti-hermitian field
\begin{equation}
	\Gamma^\mu=
	\frac{1}{2}(\xi^{\dagger}\partial^\mu\xi
	+\xi\partial^\mu\xi^{\dagger}) \;.
\end{equation}
The operator $D^\mu$ can only appear in the lagrangian in combination
with a commutator bracket, i.e. as
\begin{equation}
	[D^\mu, B] = \partial^\mu B + [\Gamma^\mu, B] \;.
\end{equation}

The chiral symmetry of the lagrangian Eq.~(\ref{lag1}) is clear,
because in the above representation {\em all\/} the fields transform
under $\text{SU(3)}_{\text{L}} \times \text{SU(3)}_{\text{R}}$ as
\begin{equation}
	X \rightarrow U X U^{\dagger}
\end{equation}
where, given
\begin{equation}
	\xi^2 \rightarrow {\xi'}^2 =
	V_{\text{L}} \xi^2 V_{\text{R}}^{\dagger}
	\text{\em\ \ with\ \ }
	V_{\text{L}}, V_{\text{R}} \in \text{SU(3)}\;,
\end{equation}
$U$ is implicitly defined by
\begin{equation}
	\xi' = V_{\text{L}} \xi U^{\dagger} =
	U \xi V_{\text{R}}^{\dagger} \;.
\end{equation}
We do not consider external fields in our formulation.

Since \cite{gass88}
\begin{equation}
	i\gamma^\mu[D_\mu,B]-\dot{m} B = {\cal O}(p) \;,
\end{equation}
where $p$ is the nucleon four-momentum, the strong interaction
lagrangian Eq.~(\ref{lag1}) counts as order ${\cal O}(p)$ in the
momentum expansion. However, there is drastic difference in the
behavior of chiral loops iterated with the lagrangian Eq.~(\ref{lag1})
and the familiar chiral perturbation theory lagrangian in the meson
sector.  Unlike in the meson sector, in the baryon sector {\em any\/}
higher-order loop starts to contribute at order ${\cal O}(p^2)$,
i.e.\ there is no correspondence between the loop expansion and the
momentum expansion \cite{gass88}. The reason for this is that the
nucleon mass in the chiral limit, $\dot{m}$, is comparable with the
chiral symmetry breaking scale $\Lambda_{\chi SB}$.  To deal with this
problem, baryon chiral perturbation theory was reformulated
\cite{jenk91}.  It starts by redefining the baryon field according to
\begin{equation}
\label{schaap}
	B_v=\text{e}^{i\dot{m} v\cdot x}B \;,
\end{equation}
where $v^\mu$ is the baryon four-velocity satisfying $v^2=1$.
Using
\begin{equation}
\label{cov_on_bb}
	iD^\mu B = \text{e}^{-i\dot{m}v\cdot x} (\dot{m}v^\mu
	+ iD^\mu )B_v
\end{equation}
the lagrangian reads, in terms of the new velocity-dependent baryon
field, $B_v$,
\begin{eqnarray}
\label{mus}
	 \text{Tr}&&\,\bar{B}_v i\gamma^\mu[D_\mu,B_v]
	-\dot{m}\bar{B}_v(1-{v\!\!\!\slash})B_v
\nonumber\\&&\mbox{}
	+D\bar{B_v}i\gamma^5\gamma^\mu\{\Delta_\mu,B_v\}
	+F\bar{B_v}i\gamma^5\gamma^\mu[\Delta_\mu,B_v] \;.
\end{eqnarray}
Next, one defines the projected fields
\begin{equation}
\label{projfields}
	B^{(+)}_{v} =  {P}^+_v B_v\;; \;\;
	B^{(-)}_{v} =  {P}^-_v B_v \;,
\end{equation}
where $P_v^+$ and $P_v^-$ are the projection operators
\begin{equation}
	 {P}^{\pm}_v= \frac{1\pm v\!\!\!\slash}{2} \;.
\end{equation}
We will show later that the minus component field, $B_v^{(-)}$, is
suppressed by $1/\dot{m}$ as compared to the plus component field,
$B_v^{(+)}$.  Therefore, in leading order in the $1/\dot{m}$ expansion
we can drop the minus component field.  Using the operators
\begin{equation}
\label{op1}
	P^+_v \gamma^\mu P^+_v =  P_v^+ v^\mu
\end{equation}
and
\begin{equation}
\label{op2}
	S_v^\mu \equiv P_v^+ \gamma_5\gamma^\mu P_v^+ \;,
\end{equation}
we find that the lowest order HBCPT lagrangian for the strong
interaction reads
\begin{eqnarray}
\label{lagleading}
	{\cal L}_{v}&=&\text{Tr}\,\bar{B}^{(+)}_{v} i v^\mu
	[D_\mu,B^{(+)}_{v}]
	-2 i D\bar{B}^{(+)}_{v} S_v^\mu \{\Delta_\mu\;,
	B^{(+)}_{v}\}
\nonumber\\&&\mbox{}
	-2 i F\bar{B}^{(+)}_{v} S_v^\mu [\Delta_\mu\;,
	B^{(+)}_{v}] \;.
\end{eqnarray}

It can be seen from Eq.~(\ref{lagleading}) that these new baryon fields
obey the modified free Dirac equation
\begin{equation}
	 i v \cdot \partial B^{(+)}_{v} = 0 \;,
\end{equation}
which no longer has the mass term. In momentum space, with $p^\mu$
the four-momentum of the baryon, and using Eq.~(\ref{schaap}), we see
that derivatives of $B^{(+)}_{v}$ produce powers of
\begin{equation}
	k^\mu = p^\mu -\dot{m} v^\mu \;,
\end{equation}
which is (for processes at low energies) a small four-momentum.
Therefore, the residual baryon momentum, $k$, is the effective
expansion parameter in this formulation of baryon chiral perturbation
theory, and the lagrangian Eq.~(\ref{lagleading}) counts as ${\cal
O}(k)$.


\subsection{Next-to-leading order}

\subsubsection{General lagrangian}

A general term in the strong interaction lagrangian of
HBCPT---considering only terms relevant to processes involving one
baryon---is of the form \begin{mathletters}
\label{genterm}
\begin{equation}
\label{genterm1}
	\text{Tr}\,\bar{H} \Gamma_v A_1 H A_2 \;,
\end{equation}
\begin{equation}
\label{genterm2}
	\text{Tr}\,\bar{H} \Gamma_v A_1
	H A_2\times\text{Tr}\,A_3 \;,
\end{equation}
or
\begin{equation}
\label{genterm3}
	\text{Tr}\,\bar{H}A_1\times\Gamma_v\times
	\text{Tr}\,A_2H\times\text{Tr}\,A_3 \;.
\end{equation}
\end{mathletters}
Here, and in the following, $H$ is defined as the ($v$-dependent) plus
component of the baryon field, i.e.
\begin{equation}
	H=B^{(+)}_{v} \;,
\end{equation}
$\Gamma_v$ can be any ($v$-dependent) operator in Dirac space,
and $A_1$, $A_2$, and $A_3$ can each be any combination of the
bosonic fields.

For the fields contained in Eq.~(\ref{genterm}) we can use $D^\mu$ and
$\Delta^\mu$, but also the hermitian scalar field $\sigma$ and the
anti-hermitian pseudoscalar field $\rho$, defined by
\begin{equation}
	\sigma=\frac{1}{2}(\xi \chi^{\dagger}\xi+\xi^{\dagger}
	\chi\xi^{\dagger});\;\;
	\rho=\frac{1}{2}(\xi \chi^{\dagger}\xi-\xi^{\dagger}
	\chi\xi^{\dagger}) \;,
\end{equation}
where $\chi$ is the chiral- and SU(3)$_{\text{F}}$ symmetry
breaking mass matrix
\begin{equation}
	\chi=B\,{\rm diag}(0,0,m_s) \;.
\end{equation}
(For simplicity we take $m_{\rm u}=m_{\rm d}=0$.)
Some combinations of $\Delta^\mu$ and $D^\mu$ are related, e.g.
one has \cite{krau90}
\begin{equation}
	[D^\mu, D^\nu] = -[\Delta^\mu, \Delta^\nu]  \;.
\end{equation}

Two of the possible operators in Dirac space have been already
given by Eqs.~(\ref{op1}) and~(\ref{op2}).
Since
\begin{equation}
	P^+_v \gamma_5 P^+_v = 0
\end{equation}
and
\begin{equation}
	P^+_v \sigma^{\mu\nu} P^+_v = -2 i [ S^\mu_v,S^\nu_v ] \;,
\end{equation}
we then find that $\Gamma_v$ in Eq.~(\ref{genterm}) can be one of
the operators
\begin{equation}
\label{gamma}
	\openone;\;\; P^+_v v^\mu;\;\; S^\mu_v;\;\;
	[S^\mu_v,S^\nu_v]
\end{equation}
combined with a general tensor constructed from $g^{\mu\nu}$,
$\epsilon^{\mu\nu\rho\lambda}$, and $v^\mu$. The latter is generated by
the covariant derivative $iD^\mu$ as in Eq.~(\ref{cov_on_bb}).  Note
that none of the terms in Eq.~(\ref{gamma}) behaves like a pseudoscalar
and $v\cdot S_v =0$ with $v^2=1$.

Unlike in the SU(2) chiral perturbation theory, where the quark mass is
usually counted as ${\cal O}(k^2)$, we have in SU(3) chiral
perturbation theory  simultaneous expansions in the independent
variables $k$ and $m_{\text{s}}$.  The leading order lagrangian
Eq.~(\ref{lagleading})  is of  order ${\cal O}(k)$.  In {\em
next-to-leading order\/} we then have all the terms of order ${\cal
O}(k^2)$ and ${\cal O}(m_{\text{s}})$.  Note that it is a matter of
definition to count the  ${\cal O}(m_{\text{s}})$ terms as
next-to-leading order.  In the heavy baryon formulation, we expect in
general to have in addition an expansion in ${\cal O}(1/\dot{m})$.
This will be discussed later.

To obtain the chiral power of a given term in the lagrangian we use
that the fields $\Delta^\mu$ and $D^\mu$ count as order ${\cal O}(k)$,
while the fields $\sigma$ and $\rho$ count as order ${\cal
O}(m_{\text{s}})$.  All the matrices in Dirac space count as order
${\cal O}(1)$.

To arrive at the next-to-leading order strong lagrangian we follow
similar lines as in Ref.~\cite{krau90}.  We will demand that a given
term in the lagrangian is Lorentz invariant, space-reversal invariant,
charge conjugation invariant, and hermitian.

\subsubsection{Terms in the lagrangian with one trace}

There are two possible terms of the form Eq.~(\ref{genterm1}), with
aside from the two baryon fields {\em one\/} additional field, A.  In
order to study the charge conjugation invariance and hermiticity it is
convenient to write these two terms in the form
\begin{equation}
\label{brackets}
	\text{Tr}\,\bar{H}\Gamma_v\{A,H\}\;;\;\;
	\text{Tr}\,\bar{H}\Gamma_v[A,H] \;.
\end{equation}
In the following we will use the short notation
\begin{equation}
	\biglb(A_1,A_2\bigrb) =
	\{A_1,A_2\} \text{ \em or } [A_1,A_2] \;
\end{equation}
for the (anti)commutator brackets,
i.e., Eq.~(\ref{brackets}) is equivalent to
\begin{equation}
\label{gt_of}
	\text{Tr}\,\bar{H}\Gamma_v\biglb(A,H\bigrb) \;.
\end{equation}

Under the charge conjugation operation the baryon field behaves
as
\begin{equation}
	H \rightarrow C \bar{H}^{\text{T}}\;,
\end{equation}
where $C$ is the charge conjugation operator, in
our representation given by $C=i\gamma^0\gamma^2$.
Since $H = P^+_v \text{e}^{i\dot{m} v\cdot x} B$,
we obtain for the charge conjugated baryon field
\begin{equation}
	H^{\text{c}} =
	P^-_v \text{e}^{i\dot{m} (-v) \cdot x} B^{\text{c}} \;,
\end{equation}
where $B^{\text{c}} = C\bar{B}^{\text{T}}$.
Therefore, the charge conjugation operation on a bilinear term in
the baryon fields reads
\begin{equation}
	\left[\bar{H} \Gamma_v H\right]^{\text{c}}
	=\bar{H}^{\text{c}} \Gamma_{-v} H^{\text{c}} \;.
\end{equation}
For the terms of the form Eq.~(\ref{gt_of}) we then have
under charge conjugation
\begin{eqnarray}
\label{cc_of}
	\left[\text{Tr}\,\bar{H}\Gamma_v\biglb(A,H\bigrb)
	\right]^{\text{c}}&=&\text{Tr}\,\bar{H}^{\text{c}}
	\Gamma_{-v}\biglb(A^{\text{c}},H^{\text{c}}\bigrb)
\nonumber\\&=&
	(-1)^{c_{\Gamma_v}+c_A}
	\text{Tr}\,\bar{H}\Gamma_v\biglb(A,H\bigrb) \;,
\end{eqnarray}
where the constants $c_{\Gamma_v}$ and $c_A$ are
implicitly defined by
\begin{equation}
\label{hond}
	 A^{\text{c}} = (-1)^{c_A} A^{\rm T};\;\;
	 C^{-1}\Gamma_{-v}C=(-1)^{c_{\Gamma_v}} \Gamma_v^{\rm T} \;.
\end{equation}
In Table~\ref{tab:fields} we display for all fields the
constants $c_A$, and in Table~\ref{tab:matrices} we display for all
operators in Dirac space the constants $c_{\Gamma_v}$.

Under complex conjugation we have for terms of the form
Eq.~(\ref{gt_of})
\begin{eqnarray}
\label{hh_of}
	\left[\text{Tr}\,\bar{H}\Gamma_v\biglb(A,H\bigrb)\right]^{*}&=&
	\text{Tr}\,\bar{H} \gamma^0\Gamma_v^\dagger\gamma^0
	\biglb(A^\dagger,H\bigrb)
\nonumber\\
	&=& (-1)^{h_{\Gamma_v}+h_A}
	\text{Tr}\,\bar{H}\Gamma_v\biglb(A,H\bigrb) \;.
\end{eqnarray}
The constants $h_A$ and $h_{\Gamma_v}$ in Eq.~(\ref{hh_of}) are defined
by
\begin{equation}
\label{hond2}
	A^{\dagger}=(-1)^{h_A}A;\;\;
	\gamma_0\Gamma_v^{\dagger}\gamma_0=
	(-1)^{h_{\Gamma_v}}\Gamma_v \;.
\end{equation}
Again, these constants can be found in Tables \ref{tab:fields}
and~\ref{tab:matrices}.

The properties of the fields and bilinear products of the Dirac
matrices under Lorentz transformations are well known (see e.g.
Ref.~\cite{krau90}), and are summarized in Tables \ref{tab:fields}
and~\ref{tab:matrices}.

{}From charge conjugation invariance and Eq.~(\ref{cc_of}) it follows
that we have terms of the form of Eq.~(\ref{gt_of}) {\em only if
$c_{\Gamma_v}+c_A$ is even}.  Hermiticity can easily be established
by using Eq.~(\ref{hh_of}) and by
multiplying a term by the complex number
$i$, if $h_{\Gamma_v}+h_A$ is odd.
Lorentz- and space-reversal invariance of terms is secured by.
contracting all the free Lorentz indices with appropriate tensors.

One readily finds that, except for the terms already encountered in the
leading order lagrangian Eq.~(\ref{lagleading}), the only allowed terms
with one field are the two SU(3) breaking terms
\begin{equation}
\label{kameel3}
	i\text{Tr}\, \bar{H} \biglb(\sigma,H\bigrb) \;.
\end{equation}
The terms in Eq.~(\ref{kameel3}) are of order ${\cal
O}(m_{\text{s}})$ and therefore belong, in our
classification scheme, to the next-to-leading order
lagrangian.

We will now consider all possible terms of the form
Eq.~(\ref{genterm1}) containing {\em two\/} bosonic fields, aside from
the baryon fields.  Closer inspection shows that the six possible
combinations can be written as
\begin{equation}
\label{neushoorn}
	\text{Tr}\,\bar{H}\Gamma_v
	\biglb({}^1A_1,\biglb({}^2 A_2,H {}^2\bigrb)
	{}^1\bigrb);\;\;\;
 	\text{Tr}\,\bar{H}\Gamma_v\biglb([A_1,A_2],H\bigrb) \;.
\end{equation}
Below it will become clear that this is a convenient way of writing the
terms down.  We have used the superscripts 1 and 2 in the first term of
Eq.~(\ref{neushoorn}) in order to make the distinction between the
(anti)commutator brackets associated with the fields $A_1$ and $A_2$.
{}From now we will assume it is clear that these brackets can be
different, and we will therefore drop the superscript. We will also
assume that in a equation where a given field appears in two terms, as
in Eq.~(\ref{cc_tf}) below, the bracket associated with that field is
the same in both terms.  Under charge conjugation the terms in
Eq.~(\ref{neushoorn}) behave as
\begin{mathletters}
\label{cc_tf}
\begin{eqnarray}
\label{cc_tf1}
	\bigl[\text{Tr}\,\bar{H}\Gamma_v \biglb(A_1,
	\biglb(A_2,H\bigrb)\bigrb)&&\bigr]^{\text{c}} =
	(-1)^{c_{\Gamma_v}+c_{A_1}+ c_{A_2}}
 \nonumber\\&&\mbox{}
	\times \text{Tr}\,
	\bar{H}\Gamma_v\biglb(A_2,\biglb(A_1,H\bigrb)\bigrb)
\end{eqnarray}
and
\begin{eqnarray}
	\bigl[\text{Tr}\,\bar{H}\Gamma_v\biglb([A_1,A_2],
	H\bigrb)&&\bigr]^{\text{c}}=
	(-1)(-1)^{c_{\Gamma_v}+c_{A_1}
	+c_{A_2}}
\nonumber\\&&\mbox{}
	\times\text{Tr}\,\bar{H}\Gamma_v\biglb([A_1,A_2],H\bigrb) \;,
\end{eqnarray}
\end{mathletters}
while under complex conjugation we have for such terms
\begin{mathletters}
\label{hh_tf}
\begin{eqnarray}
	\bigl[\text{Tr}\,\bar{H}\Gamma_v\biglb(A_1,
	\biglb(A_2,H\bigrb)\bigrb)
	&&\bigr]^{*}=
	(-1)^{h_{\Gamma_v}+h_{A_1}+ h_{A_2}}
\nonumber\\&&\mbox{}
	\times\text{Tr}\,
	\bar{H}\Gamma_v\biglb(A_2,\biglb(A_1,H\bigrb)\bigrb)
\end{eqnarray}
and
\begin{eqnarray}
	\bigl[\text{Tr}\,\bar{H}\Gamma_v
	\biglb([A_1,A_2],H\bigrb)\bigrb)&&\bigr]^*=
	(-1)(-1)^{h_{\Gamma_v}+h_{A_1}
	+h_{A_2}}
\nonumber\\&&\mbox{}
	\times\text{Tr}\,
	\bar{H}\Gamma_v\biglb([A_1,A_2],H\bigrb) \;.
\end{eqnarray}
\end{mathletters}
It can be seen from Eq.~(\ref{cc_tf}), that the allowed terms
containing two fields, $A_1$ and $A_2$, are given by
\begin{equation}
\label{spin}
	\text{Tr}\,\bar{H}\Gamma_v\biglb(A_1,
	\biglb(A_2,H\bigrb)\bigrb)+
	\text{Tr}\,\bar{H}\Gamma_v\biglb(A_2,
	\biglb(A_1,H\bigrb)\bigrb)
\end{equation}
if $c_{\Gamma_v}+c_{A_1}+c_{A_2}$ is {\em even\/},
and given by
\begin{equation}
	\text{Tr}\,\bar{H}\Gamma_v\biglb([A_1,A_2],H\bigrb)
\end{equation}
if $c_{\Gamma_v}+c_{A_1}+c_{A_2}$ is {\em odd\/}.  Hermiticity can
easily be established by using Eq.~(\ref{hh_tf}).  Combining the
building blocks of Tables \ref{tab:fields} and~\ref{tab:matrices} we
then find that the allowed terms with two fields, up to next-to-leading
order, are
\begin{mathletters}
\label{vogel}
\begin{eqnarray}
	\text{Tr}\,&&\bar{H}[v\cdot D,[v\cdot D,H]]
\\
	\text{Tr}\,&&\bar{H} [D^\mu,[D_\mu,H]]
\\
	\text{Tr}\,&&\bar{H}[S_v^\mu,
	S_v^\nu]\biglb([D_\mu,D_\nu],
	H\bigrb)
\\
\label{vogel4}
	\text{Tr}\,&&\bar{H} S_v^\mu [D_\mu,\biglb(v\cdot
	\Delta,H\bigrb)]
	+\text{Tr}\,\bar{H} S^\mu_v \biglb(v\cdot\Delta,[D_\mu,H]\bigrb)
\\
\label{vogel5}
	\text{Tr}\,&&\bar{H} S_v^\nu[v\cdot D,\biglb(\Delta_\nu,
	H\bigrb)]
	+\text{Tr}\,\bar{H} S_v^\nu\biglb(\Delta_\nu,[v\cdot D,H]\bigrb)
\\
\label{vogel6}
	\text{Tr}\,&&\bar{H} \biglb(\Delta^\mu,\biglb(
	\Delta_\mu,H\bigrb)\bigrb)
\\
\label{vogel7}
	\text{Tr}\,&&\bar{H} \biglb(v\cdot\Delta,\biglb(v\cdot
	\Delta,H\bigrb)\bigrb) \;.
\end{eqnarray}
\end{mathletters}
Since $\{A,[A,B]\}=[A,\{A,B\}]$,  only three of the four terms in Eqs.
(\ref{vogel6}) and~(\ref{vogel7}) are independent.  All the terms in
Eq.~(\ref{vogel}) are of order ${\cal O}(k^2)$ in the expansion. It can
be easily seen that the terms in Eqs. (\ref{vogel4}) and~(\ref{vogel5})
couple an {\em odd\/} number of mesons to the baryon, while the other
terms in Eq.~(\ref{vogel}) couple an {\em even\/} number.

\subsubsection{Terms in the lagrangian with more than one trace}

Now we turn to the general terms with more than one trace in flavor
space, given by Eqs. (\ref{genterm2}) and~(\ref{genterm3}).  For terms
of the form Eq.~(\ref{genterm2}) we have under charge conjugation,
\begin{eqnarray}
\label{cc_dt1}
	\bigl[\text{Tr}\,\bar{H}\Gamma_v A_1
	H A_2 \times&& \text{Tr}\,A_3\bigr]^{\text{c}}
	=
\nonumber\\&&
	\left[\text{Tr}\, \bar{H}\Gamma_v A_1 HA_2\right]^{\text{c}}
	\times \text{Tr}\,A_3^{\text{c}} \;,
\end{eqnarray}
while under complex conjugation we have
\begin{eqnarray}
\label{hh_dt1}
	\bigl[\text{Tr}\,\bar{H}\Gamma_v A_1
	H A_2 \times&& \text{Tr}\,A_3\bigr]^{\text{*}}
	=
\nonumber\\&&
	\left[\text{Tr}\,\bar{H}\Gamma_v A_1 H A_2\right]^{\text{*}}
	\times \text{Tr}\,A_3^{\dagger} \;.
\end{eqnarray}
The right-hand side of Eq.~(\ref{cc_dt1}) can be obtained using
Eqs.~(\ref{cc_of}) and~(\ref{cc_tf}), and the right-hand side of
Eq.~(\ref{hh_dt1}) can be obtained using Eqs.~(\ref{hh_of})
and~(\ref{hh_tf}).  Since
\begin{equation}
\label{tr_prop}
	\text{Tr}\,\Delta^\mu = 0\;;\;\; \text{Tr}\,[D^\mu, X] = 0
	\;\;\text{\em for any traceless field }X \;,
\end{equation}
it is easily established that, up to next-to-leading order, all
possible terms of the form Eq.~(\ref{genterm2}) are
\begin{mathletters}
\label{dbl_trace}
\begin{eqnarray}
 	\text{Tr}\,&&\bar{H}H\times\text{Tr}\,\sigma
\\
\label{dbl_trace1}
	\text{Tr}\,&&\bar{H}H\times\text{Tr} \,\Delta\cdot\Delta
\\
\label{dbl_trace2}
	\text{Tr}\,&&\bar{H}H\times\text{Tr}\,
	(v\cdot \Delta)^2 \;.
\end{eqnarray}
\end{mathletters}

For terms of the form Eq.~(\ref{genterm3}) we have under charge
conjugation and complex conjugation
\begin{eqnarray}
 	\bigl[&&\text{Tr}\,\bar{H}A_1\times\Gamma_v\times
	\text{Tr}\,A_2H\times \text{Tr}\,A_3\bigr]^{\text{c}}
	=
\nonumber\\&&
	(-1)^{c_{\Gamma_v}+c_{A_1} + c_{A_2} + c_{A_3}}
	\text{Tr}\,\bar{H}A_2 \times\Gamma_v\times\text{Tr}\,A_1H
	\times\text{Tr}\,A_3
\end{eqnarray}
and
\begin{eqnarray}
\label{dbl_trc_hh}
 	\bigl[&&\text{Tr}\,\bar{H}A_1\times\Gamma_v
	\times\text{Tr}\,A_2H
	\times\text{Tr}\, A_3\bigr]^{*} =
\nonumber\\&&
	(-1)^{h_1+h_2+h_3+h_{\Gamma_v}}
	\text{Tr}\,\bar{H}A_2\times\Gamma_v\times\text{Tr}\,A_1H
	\times\text{Tr}\, A_3\;,
\end{eqnarray}
respectively.
It then easily follows that all these are given by
\begin{mathletters}
\label{dbl_trace_twee}
\begin{eqnarray}
	\text{Tr}\,&&\bar{H}\Delta^\mu\times\text{Tr}\,\Delta_\mu H
\\
	\text{Tr}\,&&\bar{H}\Delta_\mu\times[S^\mu_v,S^\nu_v]\times
	\text{Tr}\,\Delta_\nu H
\\
	\text{Tr}\,&&\bar{H}v\cdot\Delta\times\text{Tr}\,v\cdot\Delta
	H \;.
\end{eqnarray}
\end{mathletters}

Not all the double trace terms in the lagrangian are independent due to
the matrix relation by Cayley:  For four traceless $3 \times 3$
matrices, $A$, $B$, $C$, and $D$, one has
\begin{eqnarray}
\label{cayley}
	\text{Tr}&&\bigl(DABC +  DCAB + DACB + DBAC + DBCA
\nonumber\\
	&&\mbox{}+ DCBA\bigr)
	= \text{Tr}\,DC \times \text{Tr}\,AB  +
	\text{Tr}\,DB \times \text{Tr}\,AC
\nonumber\\
	&&\mbox{}+
	\text{Tr}\,DA \times \text{Tr}\,BC \;.
\end{eqnarray}
Cayley's identity, Eq.~(\ref{cayley}), can be rewritten in a form more
suitable for our application as
\begin{eqnarray}
\label{trace_rel}
	\text{Tr}\,&&DC\times\text{Tr}\,AB =
\nonumber\\
	&&\frac{3}{4}\text{Tr}\,\bigl( D\{A,\{B,C\}\} + D\{B,\{A,C\}\}
	\bigr)
\nonumber\\
	&&+\frac{1}{4}\text{Tr}\,\bigl( D[A,[B,C]] + D[B,[A,C]]
	\bigr)
\nonumber\\
	&&\mbox{}-\text{Tr}\,DB\times\text{Tr}\,AC -
	\text{Tr}\,DA \times \text{Tr}\,BC \;.
\end{eqnarray}
Using Eq.~(\ref{trace_rel}) and the fact that $\bar{H}$, $H$, and
$\Delta^\mu$ are traceless $3 \times 3$ matrices in flavor space, one
finds
\begin{eqnarray}
 	\text{Tr}\,&&\bar{H}H\times\text{Tr}\,\Delta\cdot\Delta =
 	\frac{3}{2}\text{Tr}\,\bar{H}\{\Delta^\mu,
	\{\Delta_\mu,H\}\}
\nonumber\\
	&&\mbox{}+\frac{1}{2}\text{Tr}\,\bar{H}
	[\Delta^\mu,[\Delta_\mu,H]]
	- 2 \text{Tr}\,\bar{H}\Delta^\mu\times
	\text{Tr}\, \Delta_\mu H\;.
\end{eqnarray}
and
\begin{eqnarray}
 	&&\text{Tr}\,\bar{H}H\times\text{Tr}\,
	(v\cdot\Delta)^2  =
 	\frac{3}{2}\text{Tr}\,\bar{H}\{v \cdot \Delta,
	\{v \cdot \Delta,H\}\}
\nonumber\\
	&&\mbox{}+\frac{1}{2}\text{Tr}\,\bar{H}[v \cdot \Delta,
	[v\cdot\Delta,H]]
	- 2 \text{Tr}\,\bar{H}v \cdot\Delta \times
	\text{Tr}\,v \cdot \Delta H\;.
\nonumber\\
\end{eqnarray}
Therefore the double trace terms in Eqs.~(\ref{dbl_trace1})
and~(\ref{dbl_trace2}) are in fact a linear combination of terms in
Eqs. (\ref{vogel}) and~(\ref{dbl_trace_twee}). Note that the above
relations are also very useful for large $N_{\text{c}}$ analysis.

Counting all the independent terms in Eqs. (\ref{kameel3}),
(\ref{vogel}), (\ref{dbl_trace}), and~(\ref{dbl_trace_twee}), we find
that the the next-to-leading order lagrangian in the strong interaction
sector consists of 20 terms. Of these 20 terms, 17 terms are of order
${\cal O}(k^2)$, and 3 terms are of order ${\cal O}(m_{\text{s}})$.


\subsection{$1/\dot{m}$ corrections of the strong lagrangian}
\label{tulp}

Based on the equation of motion approach of Ref.~\cite{rho93}, we now
examine the first order correction to the HBCPT lagrangian due to the
finite nucleon mass.  To arrive at the HBCPT lagrangian
Eq.~(\ref{lagleading}) the small component of the baryon field,
$B^{(-)}_{v}$, has been dropped.  To examine this approximation
further, we go back to the lagrangian in Eq.~(\ref{mus}) obtained after
redefining the baryon field.  We first rewrite it as
\begin{equation}
\label{muis}
	\text{Tr}\,\bar{B}_v G(B_v)-\dot{m}\bar{B}_v
	(1-{v\!\!\!\slash})B_v \;,
\end{equation}
where
\begin{eqnarray}
\label{defg}
	G(B_v) &\equiv&  i\gamma^\mu[D_\mu,B_v]+D
	i\gamma^5\gamma^\mu\{\Delta_\mu,B_v\}
\nonumber\\&&\mbox{}
	+ F i\gamma^5\gamma^\mu[\Delta_\mu,B_v] \;.
\end{eqnarray}
It follows from Eq.~(\ref{muis}) that the velocity dependent baryon
field, $B_v$, satisfies the equation of motion
\begin{equation}
\label{rat}
	\dot{m} (1-{v\!\!\!\slash})B_v=G(B_v) \;.
\end{equation}
Multiplying Eq.~(\ref{rat}) on the left by the projection operator
$P_v^-$ gives
\begin{equation}
	B^{(-)}_{v}=\frac{1}{2\dot{m}}P_v^-G(B_v) \;.
\end{equation}
Since $B_v=B_v^{(-)}+B_v^{(+)}$ it follows by iteration that
\begin{equation}
\label{haas}
	B_v^{(-)}=\frac{1}{2\dot{m}}{P}^-_v
	G\left(B_v^{(+)}\right) + {\cal O}(1/\dot{m}^2) \;.
\end{equation}

Also, the equation of motion Eq.~(\ref{rat}) implies
\begin{equation}
\label{konijn}
	P_v^+G(B_v) =
	P_v^+G\left(B^{(+)}_{v} + B^{(-)}_{v}\right) = 0 \;.
\end{equation}
Substituting Eq.~(\ref{haas})
for $B^{(-)}_{v}$ in Eq.~(\ref{konijn}) leads to
\begin{equation}
	P_v^+G\left(B^{(+)}_{v}+\frac{1}{2\dot{m}}P_v^-
	G\left(B^{(+)}_{v}\right)\right)=0 \;,
\end{equation}
which can be interpreted as the equation of motion for the
$B^{(+)}_{v}$ field up to first order in $1/\dot{m}$.  We conclude from
this that the lagrangian of baryon chiral perturbation theory up to
order ${\cal O}(1/\dot{m})$ reads
\begin{equation}
	\text{Tr}\,\bar{B}^{(+)}_{v}
	G\left(B^{(+)}_{v}+\frac{1}{2\dot{m}}
	P_v^-G\left(B^{(+)}_{v}\right)\right) \;,
\end{equation}
which can be rewritten as
\begin{equation}
\label{ree}
	\text{Tr}\,\bar{B}^{(+)}_{v}
	G\left(B^{(+)}_{v}\right)
	+ \frac{1}{2\dot{m}}\text{Tr}\,\bar{B}^{(+)}_{v}
	G\left(P_v^-G\left(B^{(+)}_{v}\right)\right) \;.
\end{equation}
The first term in the right-hand side of Eq.~(\ref{ree}) corresponds to
the leading order lagrangian ${\cal L}_v$ given by
Eq.~(\ref{lagleading}), and the second term in the right-hand side of
Eq.~(\ref{ree}) is the $1/\dot{m}$ correction lagrangian
\begin{equation}
\label{roodborst}
 	{\cal L}^{1/\dot{m}} = \frac{1}{2\dot{m}}\text{Tr}\,
	\bar{B}^{(+)}_{v} G\left(P_v^-G\left(B^{(+)}_{v}\right)
	\right) \;.
\end{equation}
Using Eqs. (\ref{defg}) and~(\ref{roodborst}), and again writing
$H\equiv B_v^{(+)}$, we find that the $1/\dot{m}$ lagrangian is given
by
\begin{eqnarray}
\label{mass}
	{\cal L}&&{}^{1/\dot{m}} =
	\frac{1}{2\dot{m}}\text{Tr}\,\bar{H}\Bigl(
	[v\cdot D,[v\cdot D,H]]
\nonumber\\&&\mbox{}
	-[D^\mu,[D_\mu,H]]
	+ 2[S_v^\mu,S_v^\nu][D_\mu,[D_\nu,H]]
\nonumber\\&&\mbox{}
	-2D S_v^\mu \left( [D_\mu,\{v\cdot\Delta, H\}]+
	\{v\cdot\Delta,[D_\mu,H]\}\right)
\nonumber\\&&\mbox{}
	-2FS_v^\mu \left( [D_\mu,[v\cdot\Delta,H]]+
	[v\cdot\Delta,[D_\mu,H]]\right)
\nonumber\\&&\mbox{}
	+DF\left(\{v\cdot\Delta,[v\cdot\Delta,H]\}+
	[v\cdot \Delta,\{v\cdot D,H\}]\right)
\nonumber\\&&\mbox{}
	+D^2\{v\cdot \Delta,\{v\cdot\Delta,H\}\}
 	+F^2[v\cdot \Delta,[v\cdot\Delta,H]] \Bigr) \;.
\end{eqnarray}
Two different approaches---based on reparametrization-invariance
\cite{luke92} and using path integrals, respectively---lead to the same
result for the $1/\dot{m}$ lagrangian.

It can be seen that the terms in ${\cal L}^{1/\dot{m}}$,
Eq.~(\ref{mass}), form a subset of the terms in the general lagrangian
of order ${\cal O}(k^2)$, given in the previous section. Since the
coefficients of the higher order terms in the lagrangian will all have
to be determined by the experimental data one may question the
usefulness of calculating the $1/\dot{m}$ correction to the leading
order lagrangian.  In HBCPT there are two independent small masses, $k$
and $m_s$, and two independent large masses, $\Lambda_{\chi SB}$ and
$\dot{m}$.  In terms of extracting information from the data and then
drawing predictions from the result, one only has to control the
expansion in the small mass parameters.  It is not essential which
large masses take up the role of balancing the dimensions.  Therefore,
unless one can show that the combination of terms in Eq.~(\ref{mass})
plays a special role in the next-to-leading order corrections, it is
for a phenomenological fitting not necessary to take into account
$1/\dot{m}$ corrections from the lower order lagrangian.


\section{Weak interaction}
\label{weak}

The leading order $|\Delta{\rm S}|=1$ weak lagrangian in chiral
perturbation theory is given by \cite{dono92}
\begin{equation}
\label{lagw}
	{\cal L}_{\text{W}}=h_D\text{Tr}\,\bar{B}\{\lambda,B\}
	+h_F\text{Tr}\,\bar{B}[\lambda,B] \;,
\end{equation}
where
\begin{equation}
\label{lambda}
	\lambda\equiv\xi^{\dagger}\lambda_6\xi \;.
\end{equation}
The field $\lambda$ is constructed as in Eq.~(\ref{lambda}) in order
to make the chiral symmetry of the lagrangian Eq.~(\ref{lagw}) manifest
and to satisfy the $(8_{\text{L}},1_{\text{R}})$ transformation
property of the weak interaction.


\subsection{Leading order weak lagrangian and $1/\dot{m}$
corrections}

We now proceed in the same way as in the previous section to obtain the
weak lagrangian, including $1/\dot{m}$ corrections, in HBCPT.
Redefining the baryon field as in Eq.~(\ref{schaap}), the weak
lagrangian Eq.~(\ref{lagw}) becomes
\begin{equation}
	\text{Tr}\, \bar{B}_vG_{\text{W}}(B_v) \;,
\end{equation}
where
\begin{equation}
	G_{\text{W}}(B_v)\equiv h_D\{\lambda,B_v\}+
	h_F[\lambda,B_v] \;.
\end{equation}
Following the steps leading to Eq.~(\ref{ree}) we find that the weak
lagrangian, including the $1/\dot{m}$ correction, reads

\begin{equation}
\label{ree2}
	\text{Tr}\,\bar{H}
	G_{\text{W}}(H)
 	 + \frac{1}{2\dot{m}}
	\text{Tr}\,\bar{H}
	G_{\text{W}}\left(P_v^-G_{\text{W}}(H) \right) \;.
\end{equation}
However, in this case it can  be shown that the second term in the
right-hand side of Eq.~(\ref{ree2}) vanishes, i.e., there are {\em
no\/} $1/\dot{m}$ corrections in the weak sector, and the weak
HBCPT lagrangian simply reads
\begin{equation}
\label{stier}
	{\cal L}_{v,\text{W}} = h_D\text{Tr}\,
	\bar{H}\{\lambda,H\}
	+h_F\text{Tr}\,\bar{H}
	[\lambda,H]  \;.
\end{equation}
The $1/\dot{m}$ corrections usually are associated with only the
spin-flip part of an interaction.  The vanishing of the $1/\dot{m}$
terms can then be understood since in the weak sector the interaction
is spin independent---the structure in Dirac space is just the unity
matrix.


\subsection{The next-to-leading order weak lagrangian}

In the weak interaction both the invariance under charge conjugation,
$C$, and space-reversal, $P$, is violated.  However, to a great
accuracy the theory is invariant under their combined action, $CP$.  In
order to obtain the terms allowed in the general lagrangian, we
investigate the behavior of a given term under $CP$. Of course, we also
will demand Lorentz invariance and hermiticity. General terms in the
weak lagrangian again have the form as given by Eq.~(\ref{genterm}).

\subsubsection{Terms in the weak lagrangian with one trace}

First, we consider the case of a term with only {\em one\/} field.  All
possible terms with one field are of the form in Eq.~(\ref{gt_of}).
Under charge conjugation one has for such terms (see Eq.~(\ref{cc_of}))
\begin{eqnarray}
	\left[\text{Tr}\,\bar{H}\Gamma_v
	\biglb(A,H\bigrb)\right]^{\text{c}}&=&\text{Tr}\,
	\bar{H}^{\text{c}}\Gamma_v
	\biglb(A^{\text{c}},H^{\text{c}}\bigrb)
\nonumber\\
	&=&(-1)^{c_{\Gamma_v}}\text{Tr}\,
	\bar{H}\Gamma_v\biglb(\left(A^{\text{c}}
	\right)^{\text{T}},H\bigrb) \;,
\end{eqnarray}
where $c_{\Gamma_v}$ is defined by Eq.~(\ref{hond}).  Next, the
parity operation gives,
\begin{equation}
\label{cp_of}
	\left[\text{Tr}\, \bar{H}\Gamma_v\biglb(A,
	H\bigrb)\right]^{\text{cp}}=
	\text{Tr}\,\bar{H}\hat{\Gamma}_v\biglb(
	\hat{A},H\bigrb) \;,
\end{equation}
where we defined
\begin{equation}
\label{hatjes}
	\hat{\Gamma}_v = (-1)^{c_{\Gamma_v}} P^{-1}\Gamma_{\tilde{v}}
	P\;;\;\;
	\hat{A}=\left(\left(A^{\text{c}}\right)^{\text{T}}
	\right)^{\text{p}} \;,
\end{equation}
with $P$ the parity operator, in our representation given by
$P=\gamma^0$, and $\tilde{v}^\mu \equiv v_\mu$. Similarly, for terms
containing {\em two\/} fields, given by Eq.~(\ref{spin}), we have under
$CP$
\begin{mathletters}
\label{cp_tf}
\begin{equation}
	\left[\text{Tr}\,\bar{H}\Gamma_v\biglb(A_1,\biglb(A_2,H
	\bigrb)\bigrb)\right]^{\text{cp}}
	=\text{Tr}\,\bar{H}
	 \hat{\Gamma}_v\biglb(\hat{A}_2,
	\biglb( \hat{A}_1,H\bigrb)\bigrb)
\end{equation}
and
\begin{equation}
	\left[\text{Tr}\,\bar{H}\Gamma_v\biglb([A_1,A_2],H\bigrb)
	\right]^{\text{cp}}
	=-\text{Tr}\,\bar{H}\hat{\Gamma}_v
	\biglb([\hat{A}_1,
	\hat{A}_2],H\bigrb) \;,
\end{equation}
\end{mathletters}
respectively.  To establish hermiticity for such terms we can make use
of Eqs. (\ref{hh_of}) and~(\ref{hh_tf}).

As building blocks for terms in the general weak lagrangian we have to
our disposal the fields $\lambda$, $\Delta^\mu$, $D^\mu$, $\sigma$,
$\rho$, and the field $\lambda'$ defined by
\begin{equation}
	\lambda' = \xi^{\dagger}\lambda_7\xi \;,
\end{equation}
because $\lambda_7$ also induces s$\rightarrow$d transitions. Note that
any term must contain the field $\lambda$ or $\lambda'$, to insure the
correct $(8_{\text{L}},1_{\text{R}})$ transformation property.  For the
operators in Dirac space we may take the same as in the strong
interaction case. In Tables \ref{hatfields} and~\ref{hatmatrices} we
have displayed the necessary properties of the fields and operators
needed to apply Eqs.  (\ref{cp_of}) and (\ref{cp_tf}).  In order to
consider hermiticity we use
\begin{equation}
	\lambda^{\dagger}=\lambda;\;\;(\lambda')^{\dagger}=\lambda' \;.
\end{equation}

It easily follows that the $CP$-even terms with one bosonic field are
\begin{equation}
\label{weak_of}
	\text{Tr}\,\bar{H}\biglb(\lambda,H\bigrb) \;,
\end{equation}
i.e., indeed as in the leading order weak lagrangian ${\cal
L}_{v,\text{W}}$, Eq.~(\ref{stier}). Note that terms of the form
Eq.~(\ref{weak_of}) with $\lambda'$ instead of $\lambda$ are $CP$-odd.

Finally, the $CP$-even terms with two bosonic fields are
\begin{mathletters}
\label{single}
\begin{eqnarray}
\label{single-1}
 	i\text{Tr}\,&&\bar{H}\biglb(\lambda,[ v\cdot D,H]\bigrb) +
	i\text{Tr}\,\bar{H}[v\cdot D,\biglb(\lambda,H\bigrb)]
\\
\label{single0}
	i\text{Tr}\,&&\bar{H} S_v^\mu\biglb(\lambda,[D_\mu,H]\bigrb) +
	i\text{Tr}\,\bar{H} S_v^\mu[D_\mu,\biglb(\lambda,H\bigrb)]
\\
\label{single1}
	i\text{Tr}\,&&\bar{H}\biglb(\lambda,\biglb( v\cdot
	\Delta,H\bigrb)\bigrb)
	+i\text{Tr}\,\bar{H}\biglb( v\cdot
	 \Delta,\biglb(\lambda,H\bigrb)\bigrb)
\\
\label{single2}
	i\text{Tr}\,&&\bar{H}  S^\mu_v\biglb(\lambda,
	\biglb( \Delta_\mu,H\bigrb)\bigrb)
	+i\text{Tr}\,\bar{H} S^\mu_v\biglb(\Delta_\mu,\biglb(
	\lambda,H\bigrb)\bigrb)
\\
\label{single3}
	\text{Tr}\,&&\bar{H}\biglb([v\cdot D,\lambda'],H\bigrb)
\\
\label{single4}
	\text{Tr}\,&&\bar{H}S^\mu_v\biglb([D_\mu,\lambda'],H\bigrb)
\\
\label{single5}
	\text{Tr}\,&&\bar{H}\biglb([v\cdot \Delta,\lambda'],H\bigrb)
\\
\label{single6}
	\text{Tr}\,&&\bar{H}S^\mu_v\biglb([\Delta_\mu,\lambda'],
	H\bigrb)
\\
\label{single7}
	\text{Tr}\,&&\bar{H} \biglb(\lambda,\biglb( \sigma, H
	\bigrb)\bigrb)
	+\text{Tr}\,\bar{H} \biglb(\sigma,\biglb( \lambda, H
	\bigrb)\bigrb)
\\
\label{single8}
	i\text{Tr}\,&&\bar{H} \biglb(\lambda',\biglb( \rho, H
	\bigrb)\bigrb)
	+i\text{Tr}\,\bar{H} \biglb(\rho,\biglb( \lambda', H
	\bigrb)\bigrb)
\\
\label{single9}
	i\text{Tr}\,&&\bar{H} \biglb([\lambda',\sigma],H\bigrb)
\\
\label{single10}
	\text{Tr}\,&&\bar{H} \biglb([\lambda,\rho],H\bigrb) \;.
\end{eqnarray}
\end{mathletters}
The terms in Eq.~(\ref{single-1}) to (\ref{single6}) are of order
${\cal O}(k)$ while the terms in Eq.~(\ref{single7}) to
(\ref{single10}) are of order ${\cal O}(m_{\text{s}})$.  Note that the
parity of a given term in Eq.~(\ref{single}) is indefinite.

\subsubsection{Terms in the weak lagrangian with more
than one trace}

We now turn to the general terms in the weak interaction lagrangian
with more than one trace in flavor space, given by Eqs. (\ref{genterm2})
and~(\ref{genterm3}).  For terms of the form Eq.~(\ref{genterm2}) we
have under the combined charge conjugation and parity operation
\begin{eqnarray}
\label{cp_dt1}
	\bigl[\text{Tr}\,\bar{H}\Gamma_v A_1
	H \times&& \text{Tr}\,A_2\bigr]^{\text{cp}}
\nonumber\\&&
	= \left[\text{Tr}\, \bar{H}\Gamma_v A_1 H\right]^{\text{cp}}
	\times \left[\text{Tr}\,A_2\right]^{\text{cp}}\;,
\end{eqnarray}
which can be obtained by using Eqs. (\ref{cp_of}) and~(\ref{cp_tf}),
and
\begin{equation}
	\left[\text{Tr}\, A_3\right]^{\text{cp}}
	=\text{Tr}\,\hat{A_3} \;.
\end{equation}
For the behavior of these terms under complex conjugation we can make
use of Eq.~(\ref{hh_dt1}).  Taking into account Eq.~(\ref{tr_prop}) it
is easily established that possible $CP$-even terms of the form
Eq.~(\ref{genterm2}) in the weak sector are
\begin{mathletters}
\label{dubbel}
\begin{eqnarray}
\label{dubbel1}
 	i\text{Tr}&&\,\bar{H} H \times
	\text{Tr}\,\lambda v\cdot\Delta
\\
\label{dubbel2}
 	i\text{Tr}&&\,\bar{H}S_v^\mu H\times
	\text{Tr}\,\lambda\Delta_\mu
\\
\label{dubbel3}
 	\text{Tr}&&\,\bar{H} H \times
	\text{Tr}\,\lambda\sigma
\\
\label{dubbel4}
 	i\text{Tr}&&\,\bar{H} H \times
	\text{Tr}\,\lambda'\rho
\\
\label{dubbel5}
	\text{Tr}&&\,\bar{H}\biglb(\lambda,
	H\bigrb)\times \text{Tr}\,\sigma
\\
\label{dubbel6}
	i\text{Tr}&&\,\bar{H}\biglb(\lambda',
	H\bigrb)\times \text{Tr}\,\rho \;.
\end{eqnarray}
\end{mathletters}

For term of the form Eq.~(\ref{genterm3}) we have under a $CP$
transformation
\begin{eqnarray}
 	\bigl[\text{Tr}\,&&\bar{H}A_1\times\Gamma_v\times
	\text{Tr}\,A_2H \times \text{Tr}\,A_3\bigr]^{\text{cp}}
	=
\nonumber\\&&
	\text{Tr}\,\bar{H}\hat{A_2}
	\times\hat{\Gamma}_v\times\text{Tr}\,\hat{A_1}H
	\times\text{Tr}\,\hat{A_3}\;,
\end{eqnarray}
and for complex conjugation we can use Eq.~(\ref{dbl_trc_hh}).  It then
easily follows that all the $CP$-even terms of the form
Eq.~(\ref{genterm3}) are given by
\begin{mathletters}
\label{dubb}
\begin{eqnarray}
\label{dubb1}
	i\text{Tr}\,&&\bar{H}\lambda \times \text{Tr}\,v\cdot\Delta  H
	+i\text{Tr}\,\bar{H}v\cdot\Delta\times\text{Tr}\,\lambda H
\\
\label{dubb2}
	i\text{Tr}\,&&\bar{H}\lambda\times S_v^\mu
	\times\text{Tr}\,\Delta_\mu H
	+i\text{Tr}\,\bar{H}\Delta_\mu \times
	S_v^\mu \times \text{Tr}\,\lambda H
\\
	\text{Tr}\,&&\bar{H}\lambda' \times\text{Tr\,}v\cdot\Delta H
	-\text{Tr}\,\bar{H}v\cdot\Delta \times\text{Tr}\,\lambda' H
\\
	\text{Tr}\,\bar{H}\lambda'\times &&S_v^\mu
	\times\text{Tr}\, \Delta_\mu H
\nonumber\\
	&&\mbox{}-\text{Tr}\,\bar{H}\Delta_\mu\times S_v^\mu
	\times \text{Tr}\,\lambda' H
\\
\label{dubb3}
	\text{Tr}\,&&\bar{H}\lambda\times\text{Tr}\,\sigma H+
	\text{Tr}\,\bar{H}\sigma\times\text{Tr}\,\lambda H
\\
\label{dubb4}
	i\text{Tr}\,&&\bar{H}\lambda'\times\text{Tr}\,\rho H+
	i\text{Tr}\,\bar{H}\rho\times\text{Tr}\,\lambda' H
\\
	i\text{Tr}\,&&\bar{H}\lambda'\times\text{Tr}\,\sigma H-
	i\text{Tr}\,\bar{H}\sigma\times\text{Tr}\,\lambda' H
\\
	\text{Tr}\,&&\bar{H}\lambda\times\text{Tr}\,\rho H-
	\text{Tr}\,\bar{H}\rho\times\text{Tr}\,\lambda H \;.
\end{eqnarray}
\end{mathletters}

As in the strong interaction case,  some of the double trace terms in
the weak interaction lagrangian are related to the single trace terms.
Applying Cayley's identity, Eq.~(\ref{trace_rel}), one obtains
\begin{eqnarray}
 	i\text{Tr}\,&&\bar{H}H\times\text{Tr}\,\lambda v\cdot\Delta =
\nonumber\\&&
	\frac{3i}{4}\left(\text{Tr}\,\bar{H}\{\lambda,
	\{v\cdot\Delta,H\}\}  +
 	\text{Tr}\,\bar{H}\{v\cdot\Delta,
	\{\lambda,H\}\} \right)
\nonumber\\
	&&\mbox{}+\frac{i}{4}\left(\text{Tr}\,\bar{H}
	[\lambda,[v\cdot\Delta,H]]
	+\text{Tr}\,\bar{H}
	[v\cdot\Delta,[\lambda,H]]\right)
\nonumber\\&&\mbox{}
	-i\text{Tr}\,\bar{H}\lambda \times \text{Tr}\,v\cdot\Delta H
	-i\text{Tr}\,\bar{H}v\cdot\Delta\times\text{Tr}\,\lambda H\;,
\end{eqnarray}
i.e., the double-trace term in Eq.~(\ref{dubbel1}) can be written as a
combination of the double-trace term in Eq. (\ref{dubb1}) and two of
the single-trace terms in Eq.~(\ref{single1}).  In the same way we have
for the term in Eq.~(\ref{dubbel2})
\begin{eqnarray}
 	i&&\text{Tr}\,\bar{H}S_v^\mu H\times\text{Tr}\,
	\lambda\Delta_\mu=
\nonumber\\&&
	\frac{3i}{4}\left(\text{Tr}\,\bar{H}S_v^\mu\{\lambda,
	\{\Delta_\mu,H\}\}  +
 	\text{Tr}\,\bar{H}S_v^\mu\{\Delta_\mu,
	\{\lambda,H\}\} \right)
\nonumber\\
	&&\mbox{}+\frac{i}{4}\left(\text{Tr}\,\bar{H}S_v^\mu
	[\lambda,[\Delta_\mu,H]]
	+\text{Tr}\,\bar{H}S_v^\mu
	[\Delta_\mu,[\lambda,H]]\right)
\nonumber\\&&\mbox{}
	-i\text{Tr}\,\bar{H}\lambda
	\times S_v^\mu\times\text{Tr}\,\Delta_\mu H
	-i\text{Tr}\,\bar{H}\Delta_\mu\times S_v^\mu\times
	\text{Tr}\,\lambda H\;.
\nonumber\\
\end{eqnarray}

Since $\sigma$ and $\rho$ are non-traceless matrices in flavor space,
we cannot apply Cayley's identity for the double-trace terms in Eqs.
(\ref{dubbel3}) and~(\ref{dubbel4}). However, it can be easily shown
that for three $3\times 3$ traceless matrices $A$, $C$, and $D$, and
one arbitrary $3\times 3$ matrix $B$, one has
\begin{eqnarray}
\label{trace_rel2}
	\text{Tr}\,&&DC\times\text{Tr}\,AB =
\nonumber\\
	&&\frac{3}{4}\text{Tr}\,\bigl( D\{A,\{B,C\}\} + D\{B,\{A,C\}\}
	\bigr)
\nonumber\\
	&&+\frac{1}{4}\text{Tr}\,\bigl( D[A,[B,C]] + D[B,[A,C]]
	\bigr)
\nonumber\\
	&&\mbox{}-\text{Tr}\,DB\times\text{Tr}\,AC -
	\text{Tr}\,DA \times \text{Tr}\,BC
\nonumber\\
	&&\mbox{}-\text{Tr}\, D \{ A,C\} \times\text{Tr}\,B \;.
\end{eqnarray}
Using Eq.~(\ref{trace_rel2}) we then find that the terms in Eqs.
(\ref{dubbel3}) and~(\ref{dubbel4}) satisfy
\begin{eqnarray}
 	\text{Tr}\,&&\bar{H}H\times\text{Tr}\,\lambda \sigma =
\nonumber\\&&
	\frac{3}{4}\left(\text{Tr}\,\bar{H}\{\lambda,
	\{\sigma,H\}\}  +
 	\text{Tr}\,\bar{H}\{\sigma,
	\{\lambda,H\}\} \right)
\nonumber\\
	&&\mbox{}+\frac{1}{4}\left(\text{Tr}\,\bar{H}
	[\lambda,[\sigma,H]]
	+\text{Tr}\,\bar{H}
	[\sigma,[\lambda,H]]\right)
\nonumber\\&&\mbox{}
	-\text{Tr}\,\bar{H}\lambda \times \text{Tr}\,\sigma H
	-\text{Tr}\,\bar{H}\sigma\times\text{Tr}\,\lambda H
\nonumber\\&&\mbox{}
	-\text{Tr}\,\bar{H}\{\lambda,H\}\times\text{Tr}\,
	\sigma \;,
\end{eqnarray}
and
\begin{eqnarray}
 	i\text{Tr}\,&&\bar{H}H\times\text{Tr}\,\lambda' \rho =
\nonumber\\&&
	\frac{3i}{4}\left(\text{Tr}\,\bar{H}\{\lambda',
	\{\rho,H\}\}  +
 	\text{Tr}\,\bar{H}\{\rho,
	\{\lambda',H\}\} \right)
\nonumber\\
	&&\mbox{}+\frac{i}{4}\left(\text{Tr}\,\bar{H}
	[\lambda',[\rho,H]]
	+\text{Tr}\,\bar{H}
	[\rho,[\lambda',H]]\right)
\nonumber\\&&\mbox{}
	-i\text{Tr}\,\bar{H}\lambda' \times \text{Tr}\,\rho H
	-i\text{Tr}\,\bar{H}\rho\times\text{Tr}\,\lambda' H
\nonumber\\&&\mbox{}
	-i\text{Tr}\,\bar{H}\{\lambda',H\}\times\text{Tr}\,
	\rho \;,
\end{eqnarray}
respectively, i.e., they can both be written as a linear combination
of terms in Eqs. (\ref{single}), (\ref{dubbel}), and~(\ref{dubb}).

In total we find in the next-to-leading order $CP$-even
weak interaction lagrangian 44 independent terms, of which 24 are of
order ${\cal O}(k)$ and 20 of order ${\cal O}(m_{\text{s}})$.
The $CP$-odd lagrangian can be easily obtained from the $CP$-even
lagrangian by exchanging in every term $\lambda$ by
$\lambda'$.


\section{Summary and conclusions}
\label{conclusions}

In this paper we have considered the general lagrangian of baryon
chiral perturbation theory in the  heavy baryon formulation for the
strong- and weak interaction sector up to next-to-leading order, i.e.,
the first order correction to the leading order.  Using general
symmetry principles we gave the restrictions on a given term in the
lagrangian.  We investigated the relation between the terms with a
single and a double trace in flavor space.  In the strong interaction
sector we found a total number of 20 terms in the next-to-leading order
lagrangian, while that number in the weak interaction sector is 44. In
the weak interaction sector we only gave explicitly the $CP$-even
part; the extension to the $CP$-odd part is trivial.

We have also examined the $1/\dot{m}$ corrections.  In the weak
interaction sector these are shown to be absent.  This is due to the
simple structure of the weak interaction in Dirac space.  The
significance of this observation is unclear yet. In the strong
interaction sector the $1/\dot{m}$ lagrangian is a linear combination
of terms in the next-to-leading order lagrangian. For this reason we
argued that it is in fact not necessary to take $1/\dot{m}$ corrections
into account when doing a phenomenological fitting.

For a consistent study of low-energy processes with baryons,
e.g.\ hyperon decay, it is necessary to include all the terms in the
lagrangian in the calculation.  Starting with this lagrangian, one can
then fix by experimental data as many of the coefficients associated
with the terms in the lagrangian as possible.  Once the coefficients
are fixed, new predictions can be drawn.  This work is in progress.
Also, we have considered only the octet baryons so far.  In general,
the decuplet may also play a important role in some of the
phenomenological issues.  However, inclusion of the decuplet involves
some complications which will be resolved later.

\acknowledgments

This work is supported by the National Science Council of the ROC under
contract number NSC84-2811-M008-001, NSC84-2112-M007-042,
NSC84-2732-M008-001, and NSC84-2112-M008-013.

%
%

\begin{table}
\caption{Properties of the fields, needed for building general
terms in the baryon chiral lagrangian in the strong interaction
sector. The constants $c_A$ and $h_A$, in the second and third column
respectively, are defined by
Eqs.~(\protect\ref{hond}) and~(\protect\ref{hond2}), respectively. The
fourth column gives, in the
usual notation, the properties under the Lorentz and parity
transformation.}
\label{tab:fields}
\begin{tabular}{ccccc}
	$A$ & $c_A$ & $h_A$ & L   \\
	\hline
	$\Delta^\mu$ & 0 & 1 & A  \\
	$D^\mu$ & 1 & 1 & V \\
	$\sigma$ & 0 & 0 & S \\
	$\rho$ & 0 & 1 & PS
\end{tabular}
\end{table}

\begin{table}
\caption{ Same as Table~\protect\ref{tab:fields} for the operators
in Dirac space. They all count as order ${\cal O}(1)$ in the chiral
expansion. We only write down the operators with at most two
Lorentz indices, since operators with three or more indices will
contribute only to higher order terms.}
\label{tab:matrices}
\begin{tabular}{ccccc}
	$\Gamma_v$ & $c_{\Gamma_v}$ & $h_{\Gamma}$ &
	L \tablenote{It is here assumed that the matrices
	are in combination with the bilinear
	product of the baryon fields}\\
	\hline
	$\openone$ & 0 & 0 & S \\
	$P_v^+ v^\mu$&1&0&V\\
	$S^\mu_v$ & 0 & 0 & A\\
	$[S^\mu_v, S^\nu_v]$ & 1 & 1& T\\
	$P^+_v v^\mu v^\nu$ & 0 & 0 & T \\
	$S^\mu_v v^\nu$ & 1 & 0 & PT
\end{tabular}
\end{table}

\begin{table}
\caption{Properties of the fields, needed to consider the combined
operation of charge conjugation and parity on a general term in the
weak lagrangian. Given a field $A$, the field $\hat{A}$ is defined
by Eq.~(\protect\ref{hatjes}).}
\label{hatfields}
\begin{tabular}{c|c}
	$A$ & $\hat{A}$ \\
	\hline
	$\lambda$ & $\lambda$ \\
	$\lambda'$ & $-\lambda'$ \\
	$\Delta^\mu$ & $-\Delta_\mu$ \\
	$D^\mu$ & $-D_\mu$ \\
	$\sigma$ & $\sigma$ \\
	$\rho$ & $-\rho$
\end{tabular}
\end{table}

\begin{table}
\caption{As Table~\protect\ref{hatfields} for the operators in
Dirac space.
Again, given a operator $\Gamma_v$, the fields $\hat{\Gamma}_v$ is
defined by Eq.~(\protect\ref{hatjes}). No operators with more than one
Lorentz index are needed, since they will only appear in terms of
higher order in the weak lagrangian.}
\label{hatmatrices}
\begin{tabular}{c|c}
	$\Gamma_v$ & $\hat{\Gamma}_v$\\
	\hline
	$v^\mu$ & $-v_\mu$\\
	$S_v^\mu$ & $-(S_v)_\mu$
\end{tabular}
\end{table}

\end{document}